\documentclass[aps,superscriptaddress,showpacs,preprintnumbers,nofootinbib,amsmath,amssymb,]{revtex4}
\usepackage{braket}
\usepackage{bm}
\usepackage{dcolumn}
\usepackage{hyperref}
\usepackage[dvipdfmx,final]{graphicx}
\usepackage{ulem}
\input{colordvi.tex}
\usepackage{subfigure}
\usepackage{verbatim}
\usepackage{makeidx}
\usepackage{amssymb}
\usepackage{float}
\usepackage{url}
\usepackage{color}
\newcommand{\be}{\begin{eqnarray}}
\newcommand{\ee}{\end{eqnarray}}

\newcommand{\eV}{\,{\rm eV}}

\begin{document}

\title{Cosmological 21cm line observations  to test scenarios of
  super-Eddington accretion on to black holes being seeds of
  high-redshifted supermassive black holes }

\author{Kazunori Kohri}
\email{kohri@post.kek.jp}
\affiliation{Theory Center, IPNS, KEK, 1-1 Oho, Tsukuba, Ibaraki 305-0801, Japan}
\affiliation{The Graduate University for Advanced Studies (SOKENDAI), 1-1 Oho, Tsukuba, Ibaraki 305-0801, Japan}
\affiliation{Kavli IPMU (WPI), UTIAS, The University of Tokyo, Kashiwa, Chiba 277-8583, Japan}

\author{Toyokazu Sekiguchi}
\affiliation{Theory Center, IPNS, KEK, 1-1 Oho, Tsukuba, Ibaraki 305-0801, Japan}

\author{Sai Wang}
\affiliation{Theoretical Physics Division, Institute of High Energy Physics, Chinese Academy of Sciences, Beijing 100049, P. R. China}
\affiliation{School of Physical Sciences, University of Chinese Academy of Sciences, Beijing 100049, P. R. China}

\date{\today}

\begin{abstract} 
 In this paper, we study scenarios of the super-Eddington accretion onto black
holes at high redshifts $z > 10$, which are expected to be seeds to evolve to
supermassive black holes until redshift $z \sim 7$. For an initial mass,
$M_{\rm BH, ini} \lesssim 2 \times 10^{3} M_{\odot}$ of a seed BH, we
definitely need the super-Eddington accretion, which can be applicable to both
astrophysical and primordial origins. Such an accretion disk inevitably emitted
high-energy photons which had heated the cosmological plasma of the
inter-galactic medium continuously from high redshifts. In this case, the
cosmic history of cosmological gas temperature is modified, by which the
absorption feature of the cosmological 21 cm lines is suppressed. By comparing
theoretical predictions of the 21cm line absorption with the observational data
at $z\sim17$, we obtain a cosmological upper bound on the mass-accretion rate
as a function of the seed BH masses. In order to realize $M_{\rm BH} \sim 10^9
M_{\odot}$ at $z \sim 7$ by a continuous mass-accretion on to a seed BH, to be
consistent with the cosmological 21cm line absorption at $z \sim 17$, we
obtained an severe upper bound on the initial mass of the seed BH to be $M_{\rm
BH, ini} \lesssim 10^2 M_{\odot}$ ($M_{\rm BH, ini} \lesssim 10^6 M_{\odot}$)
when we assume a seed BH with its comoving number density $n_{\rm seed,0} \sim
10^{-3} {\rm Mpc}^{-3}$ ($n_{\rm seed,0} \sim 10^{-7} {\rm Mpc}^{-3}$). We also
discuss some implications for application to primordial black holes as the seed
black holes.
\end{abstract}

\preprint{
KEK-TH-2389\quad
KEK-Cosmo-0284\quad
}
\maketitle

\section{Introduction}
\label{sec:introduction}
Quite recently a luminous quasar (QSO) J0313-1806 was observed at redshift $z$= 7.642~\cite{Wang:2021apjl}. The mass of the central black hole (BH) in this QSO system is quite large $M_{\rm BH} = (1.6 \pm 0.4) \times 10^9 M_{\odot}$. In addition to this BH, so far several QSOs at around $z \sim $7 have been already reported, each of which has a massive central BH similarly with the order of $M_{\rm BH} \sim {\cal O}(10^9)~M_{\odot}$.  In astronomy and
astrophysics, it is a big challenge to produce such supermassive black holes (SMBHs) in an early Universe within cosmic time $t(z \sim 7) \sim 0.76$~Gyr.~\footnote{Here we used the Hubble
  parameter $H_0 = 68 {\rm km/s/Mpc}$, the omega parameters of matter,
  $\Omega_M = 0.32$ and cosmological constant
  $\Omega_{\Lambda} = 0.68$ as reference values.}

One of the most natural scenarios to increase a BH mass to a massive one should be a gas accretion on to it. However, even if the Eddington accretion is realized, there is no easy solution. For example, when we assume the Eddington accretion rate is successively-continuing with a radiative efficiency $\eta_{\rm eff} \sim 0.1$ from cosmic time $t(z=30) = 0.10$~Gyr~\cite{Barkana:2000fd} to $t(z=7) = 0.76$~Gyr, the seed mass at $z \gtrsim 30$ should be $\sim {\cal O}(10^{3.3})~M_{\odot}$ in order to obtain $M_{\rm BH} \sim {\cal O}(10^9)~M_{\odot}$ until $z=7$.  Therefore, even if we assume the Eddington accretion rate, we need a massive seed as an initial condition~\cite{Woods:2018lty,Inayoshi:2019fun}. Possible scenarios to produce such massive seed BHs have been studied, e.g., through collapses of massive stars/gas clouds, or through mergers of massive stars/ black holes~\cite{Loeb:1994wv,Omukai:2000ic,Oh:2001ex,Volonteri:2005fj,Lodato:2006hw,Wise:2007bf,Regan:2008rv,Shang:2009ij,Volonteri:2010wz,Hosokawa:2012uq,Inayoshi:2012zi,Latif:2013pyq,Hosokawa:2013mba,Regan:2014maa,Inayoshi:2014rda,Ferrara:2014wua,Becerra:2014xea,Latif:2015eoa,Chon:2016nmh,wise:2019nature,Becerra:2018mnras,Maio:2018sfz,Mayer:2017nature,Mayer:2018gyh,Mayer:2014nva,Dijkstra:2014mnras,Sugimura:2014sqa,Wolcott-Green:2016grm,Wolcott-Green:2020avn,Chon:2018mnras,Matsukoba:2018qxr,Bromm:2002hb,Spaans:2006ur,
  Sanders:1970,PortegiesZwart:2002iks,PortegiesZwart:2004ggg,Freitag:2005yd,Omukai:2008wv,Devecchi:2009,Devecchi:2012nw,katz:2015,Sakurai:2017opi,Stone:2016ryd,Reinoso:2018bfv,Tagawa:2019vep,Boekholt:2018gbw,yajima:2016mnras,Davies:2011pd,Lupi:2015wxp}. Alternatively, those massive seeds may be formed through other primordial origins such as primordial black holes~\cite{Kawasaki:2012kn,Kohri:2014lza,Nakama:2016kfq,Serpico:2020ehh,Unal:2020mts} (See also Refs.~\cite{Carr:2009jm,Green:2020jor,Carr:2021bzv,Carr:2020gox} for review articles). Another possibility would be assuming a super-Eddington accretion rate~\footnote{ For concrete models of the super-Eddington accretion, see references for the slim disk
  models~\cite{Watarai:2000,Watarai:2003tr,Yuan:2014gma}, the
  neutrino-dominated accretion flows~\cite{Kohri:2002kz,Kohri:2005tq},
  etc. and references therein.  }  on to a light seed
BH~\cite{Begelman:1978,Sadowski:2009gg,Wyithe:2012,Madau:2014pta,Inayoshi:2015pox,Pezzulli:2016,Pezzulli:2017ikf,Begelman:2016gle,Alexander:2014,Pacucci:2017mcu,Mayer:2018gyh,Takeo:2020vmm,natarajan:2021}.~\footnote{See
  also Ref.~\cite{Ebisuzaki:2001qm} for another mechanism through
  mergers of seed BHs which were produced by mergers of massive stars.}

To test consistencies of the theoretical disk models for the (super-)Eddington accretions with observations, in principle we can use cosmological 21cm line emissions/absorptions which were produced in the dark ages at $z \sim {\cal O}(10) - {\cal O}(10^2)$ (e.g., see \cite{Tanaka:2015sba,Ewall-Wice:2018bzf,Sazonov:2018tnj,Ewall-Wice:2019may,Ma:2021pgp} for earlier works).  So far, the EDGES collaboration has reported the observational data for the absorption feature of the cosmological global 21cm line spectrum at around $z\sim 17$~\cite{Bowman:2018yin}.  If there existed an extra heating source due to emissions from the accretion disks in this epoch, gas temperature could be larger than the standard value predicted in the standard $\Lambda$-cold dark matter ($\Lambda$CDM) model. In this case, a depth of the 21cm line absorption should have become shallower than the one in the $\Lambda$CDM model. Because the EDGES collaboration reported the large depth with finite errors, we can test this kind of scenarios and obtain a conservative upper bound on such an extra emission at least not to bury the observed absorption trough~\cite{DAmico:2018sxd,Hiroshima:2021bxn}.  It is known that a cosmological extra heating of the order of ${\cal O}(10^{-20}) \mathrm{eV} / \mathrm{sec} / \mathrm{cm}^{3}$ at $z\sim 17$ affects the absorption feature of the global 21cm line spectrum (e.g., see Ref.~\cite{Hiroshima:2021bxn} and references therein). In this paper, by using this logic, we discuss how we can constrain the scenarios of the (super-)Eddington accretions onto high-redshifted seed BHs which are expected to evolve to the SMBHs until $z \sim 7$ and obtain observational bounds on the accretion rates.

This paper is organized as follows.  In Sec.~\ref{sec:igm}, we review how energy injection affects evolution of the intergalactic medium (IGM).  Models of accretion disks are introduced in Sec.~\ref{sec:modelAdisk}. In Sec.~\ref{sec:injection}, we discuss energy injections from accretions onto seed black holes.  In Sec.~\ref{sec:results}, we present our results. We conclude in the final section~\ref{sec:conclusion}.

Throughout the paper, we adopt the Heaviside–Lorentz units of $c = \hbar = k_B = 1$ unless otherwise stated.

\section{Evolution equations of IGM in the presence of energy injection} 
\label{sec:igm}
In this section, we explain how energy injection affects evolutions of the IGM.  For illustrative purposes, we follow a simple description of hydrogen in the IGM, which is based on the effective three-level atom~\cite{Peebles:1968ja,Zeldovich:1969en,Seager:1999km}. As will be shown in~Sec~\ref{sec:results} however, we actually execute the public recombination code {\tt
  HyRec}\footnote{\url{https://pages.jh.edu/~yalihai1/hyrec/hyrec.html}},
which is based on the state-of-art effective multi-level atom (See \cite{AliHaimoud:2010dx,Chluba:2010ca}). In this study, for simplicity we focus only on ionization and recombination of hydrogen while assuming helium is neutral. It is known that this simplification is a good approximation as long as we are interested in processes that occurred in the cosmic dark ages (Dark Ages)~\cite{Liu:2016cnk} (See also~\cite{Liu:2019bbm}).

The cosmic evolution of the ionization fraction, $x_e$, is then described by the following equation:
\begin{eqnarray}
\frac{dx_e}{dt}&=&-C_{\rm P}\left[\alpha_{\rm H}(T_m)x_e^2 n_H-\beta_{\rm H}(T_\gamma)(1-x_e)
e^{-E_\alpha/T_\gamma}\right]\notag\\
&&\quad+\frac{dE_{\rm inj}}{dVdt}\frac1{n_{\rm H}}\left[
\frac{f_{\rm ion}(t)}{E_0}+\frac{(1-C_{\rm P})f_{\rm exc}(t)}{E_\alpha}
\right], \label{eq:dotxe}
\end{eqnarray}
where $T_m$ and $T_\gamma$ are the temperatures of gas and photon, respectively.  $n_{\rm H}$ is the number density of hydrogen, $E_0\simeq13.6\eV$ is the ionization energy of hydrogen, and $E_\alpha=3E_0/4$ is the energy of Ly-$\alpha$. Here $\alpha_{\rm H}$ is the case-B recombination coefficient, and $\beta_{\rm H}$ is the corresponding ionization rate.  The Peebles' $C$-factor ($C_{\rm P}$), represents the probability that a hydrogen atom initially in the $n=2$ shell reaches the ground state without being photoionized. It is given by
\begin{eqnarray}
  \label{eq:PeeblesC}
C_{\rm P}=\frac{\Lambda n_{\rm H}(1-x_e)+\frac1{2\pi^2}E_\alpha^3 H(t)}
{\Lambda n_{\rm H}(1-x_e)+\frac1{2\pi^2}E_\alpha^3 H(t)+\beta_Hn_H(1-x_e)},
\end{eqnarray}
where $\Lambda\simeq8.23$s$^{-1}$ is the two-photon decay rate of the hydrogen 2$s$-state, and $H(t)$ is the Hubble expansion rate. The last term in \eqref{eq:dotxe} represents the effects of energy injection with the energy injection rate per unit volume per time, $dE_{\rm inj}/(dVdt)$, which will be shown in detail in the next section.

Evolutions of the gas temperature $T_m$ is described by the following equation:
\begin{eqnarray}
\frac{dT_m}{dt} &=&
-2H(t)T_m+\Gamma_C (T_{\gamma}-T_m)
+\frac{dE_{\rm inj}}{dVdt}\frac1{n_{\rm H}}
\frac{2f_{\rm heat}(z)}{3(1+x_e+f_{\rm He})}, \label{eq:dotTm}
\end{eqnarray}
where  $\Gamma_C$ is the coupling rate of $T_m$ to $T_\gamma$, which is dominated by the Compton scattering, 
\begin{eqnarray}
  \label{eq:GammaC}
\Gamma_C=\frac{8\sigma_Ta_rT_\gamma^4}{3m_e} \frac{x_e}{1+f_{\rm He}+x_e},
\end{eqnarray}
where $\sigma_T$ is the Thomson scattering cross section, $a_r$ is radiation constant, $m_e$ is the electron mass, and $f_{\rm He}$ is the number ratio of helium to hydrogen. The last term in \eqref{eq:dotTm} represents heating of the gas temperature due to the energy injection.  As defined in~\cite{Slatyer:2015jla,Slatyer:2015kla} the coefficients $f_{\rm ion}(t)$, $f_{\rm exc}(t)$, and $f_{\rm heat}(t)$ (collectively denoted by $\{f_c(t)\}$ hereafter) are the fractions of injected energy deposited into the hydrogen ionization, the hydrogen excitation, and the heating of gas, respectively.

\begin{figure}[ht]
\centering
\includegraphics[width=10cm]{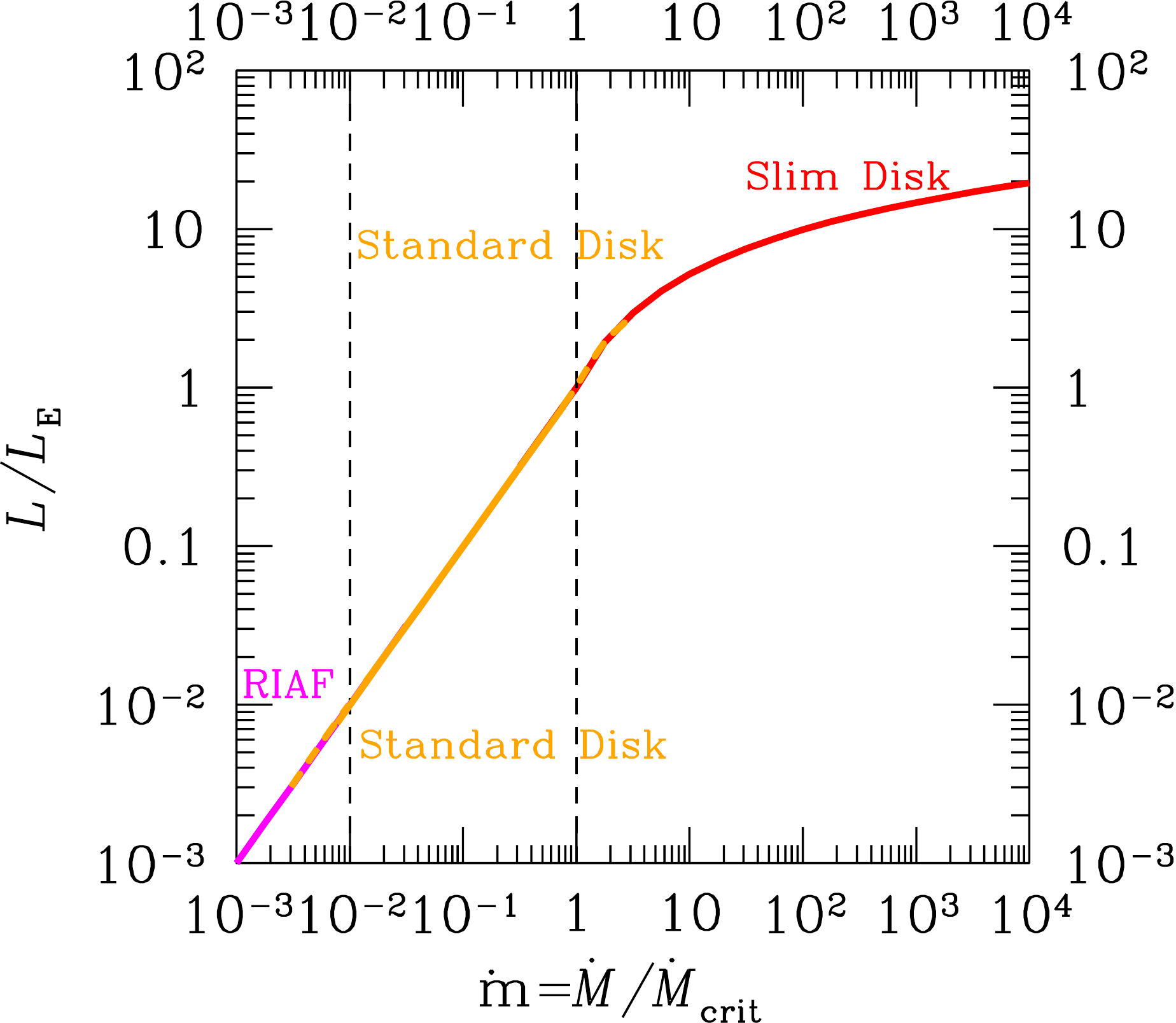}
\caption{Luminosity $L$ of the accretion disk as a function of the   accretion rates for three models, RIAF (Radiation Inefficient Accretion Flow)~\cite{Yuan:2014gma}, the   standard disk,  and the slim disk~\cite{Watarai:2000}. Here  $L_{\rm E}$ is the Eddington luminosity and $\dot{M}_{\rm crit}$ is   the corresponding critical accretion rate. In the theoretical  calculations, the dimensionless viscous parameter $\alpha_{\rm vis}$  is taken to be 0.1.}
\label{fig:lm}
\end{figure}

\section{Models of accretion disks} 
\label{sec:modelAdisk}

\begin{figure}[ht]
\centering
\includegraphics[width=10cm]{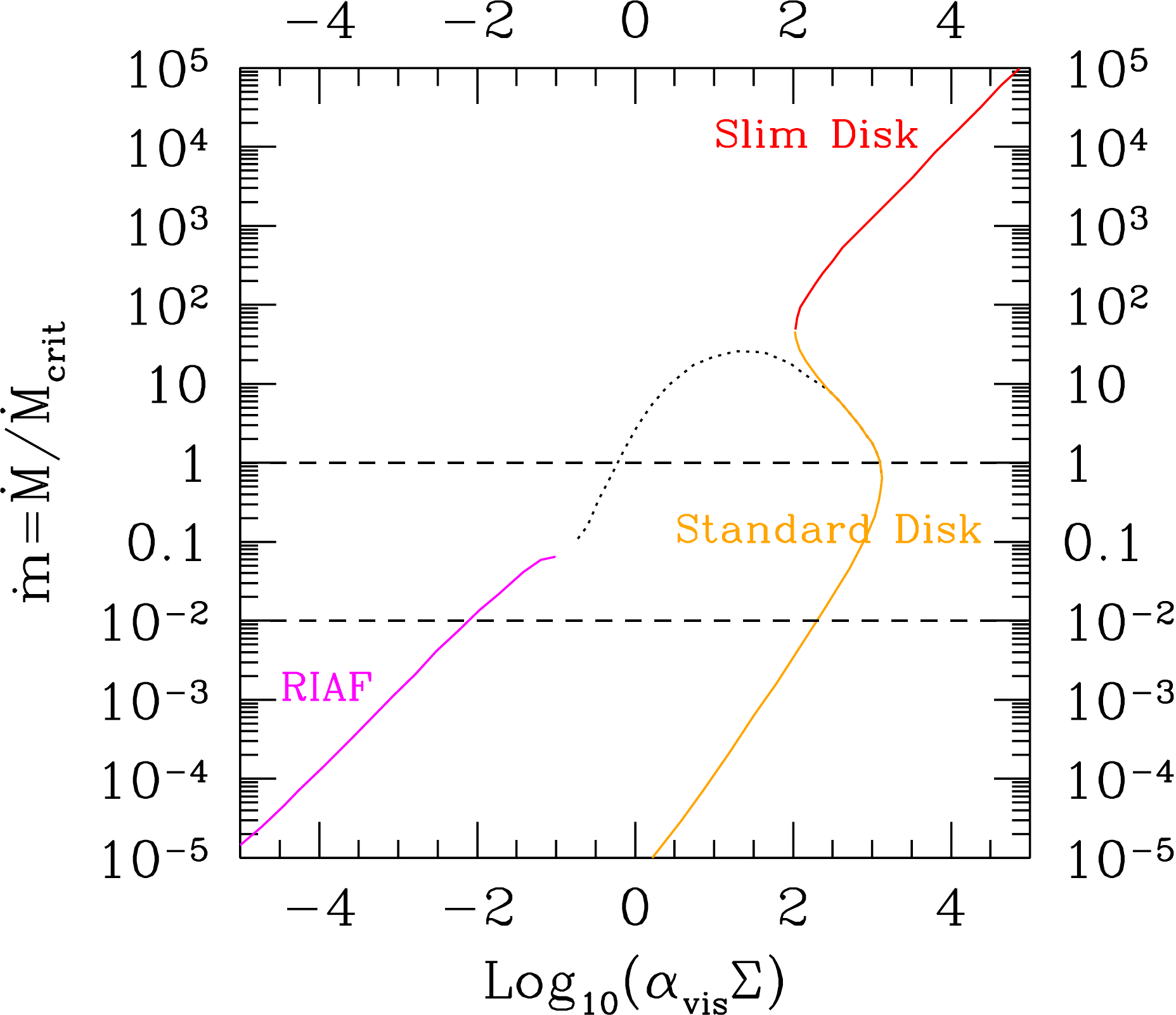}
\caption{A unified picture of accretion disk
  models~\cite{Yuan:2014gma,Kohri:2002kz}. Here we plot accretion rates as a function of $\alpha_{\rm vis} \Sigma$ for the RIAF~\cite{Yuan:2014gma}, the standard disk, and the slim disk~\cite{Watarai:2000}, respectively. Here $\Sigma = \int \rho dz$ is the surface density which is obtained by integrating the mass density $\rho$ along the vertical axis (z) with respect to the disk plane.}
\label{fig:diskModels}
\end{figure}

In Fig.~\ref{fig:lm}, we plot the luminosity $L$ from the accretion disks as a function of the accretion rate $\dot{M}$ in case of the viscous parameter $\alpha_{\rm vis}$ (=0.1) in the unified picture of the three models (see Fig.~\ref{fig:diskModels})~\cite{Yuan:2014gma}, RIAF (Radiation Inefficient Accretion Flow) ($\dot{m} \lesssim 10^{-2}$), the standard disk ($10^{-2} \lesssim \dot{m} \lesssim 1$)), and the slim disk ($1\lesssim \dot{m}$). The luminosity is normalized by the Eddington luminosity,
\begin{eqnarray}
  \label{eq:EddingtonL}
L_E 
\simeq 1.3 \times 10^{38}  \mathrm{erg \ sec^{-1}}
\left(\frac{M_{\rm BH}}{M_{\odot}}\right),
\end{eqnarray}
where $M_{\odot}$ denotes  Solar mass $(\simeq2.0 \times 10^{33}$~g).
The accretion rate is normalized by the critical accretion rate, $\dot{M}_{\rm crit}$ to be
\begin{eqnarray}
  \label{eq:smallmdot}
  \dot{m} = \frac{\dot{M}}{\dot{M}_{\rm crit}},
\end{eqnarray}
where $\dot{M}_{\rm crit} \equiv \eta_{\rm eff}^{-1} L_E$ with the radiative efficiently $\eta_{\rm eff}$, which ranges $\eta_{\rm eff}^{-1} \sim 10 - 16$ in the standard disk and the slim disk models~\cite{Watarai:2000,Mineshige:2008}, and approximately in the RIAF model with $\dot{m} \gtrsim 10^{-3}$~\cite{Yuan:2014gma}. In this study, for simplicity we take $\eta_{\rm eff}^{-1} = 10$ as a reference value. Then the critical accretion rate is represented by
\begin{eqnarray}
  \label{eq:MdotCrit}
\dot{M}_{\rm crit} \simeq 1.4 \times 10^{18} {\rm g} \ {\rm sec} ^{-1} 
\left( \frac{\eta_{\rm eff}^{-1}}{10}\right)
\left(\frac{M_{\rm BH}}{M_{\odot}}\right).
\end{eqnarray}

\begin{figure}[ht]
\centering
\includegraphics[width=8cm]{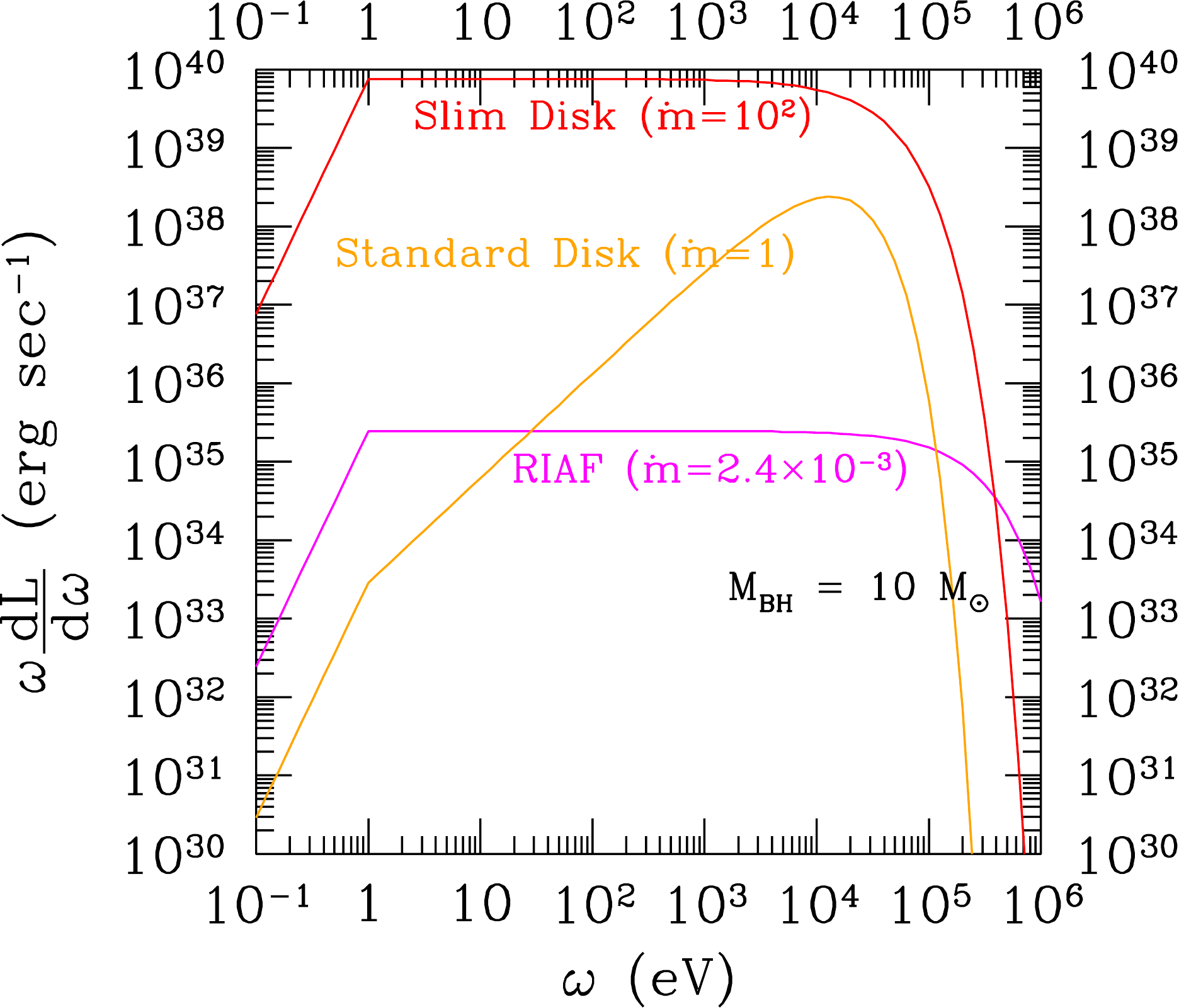}
\includegraphics[width=8cm]{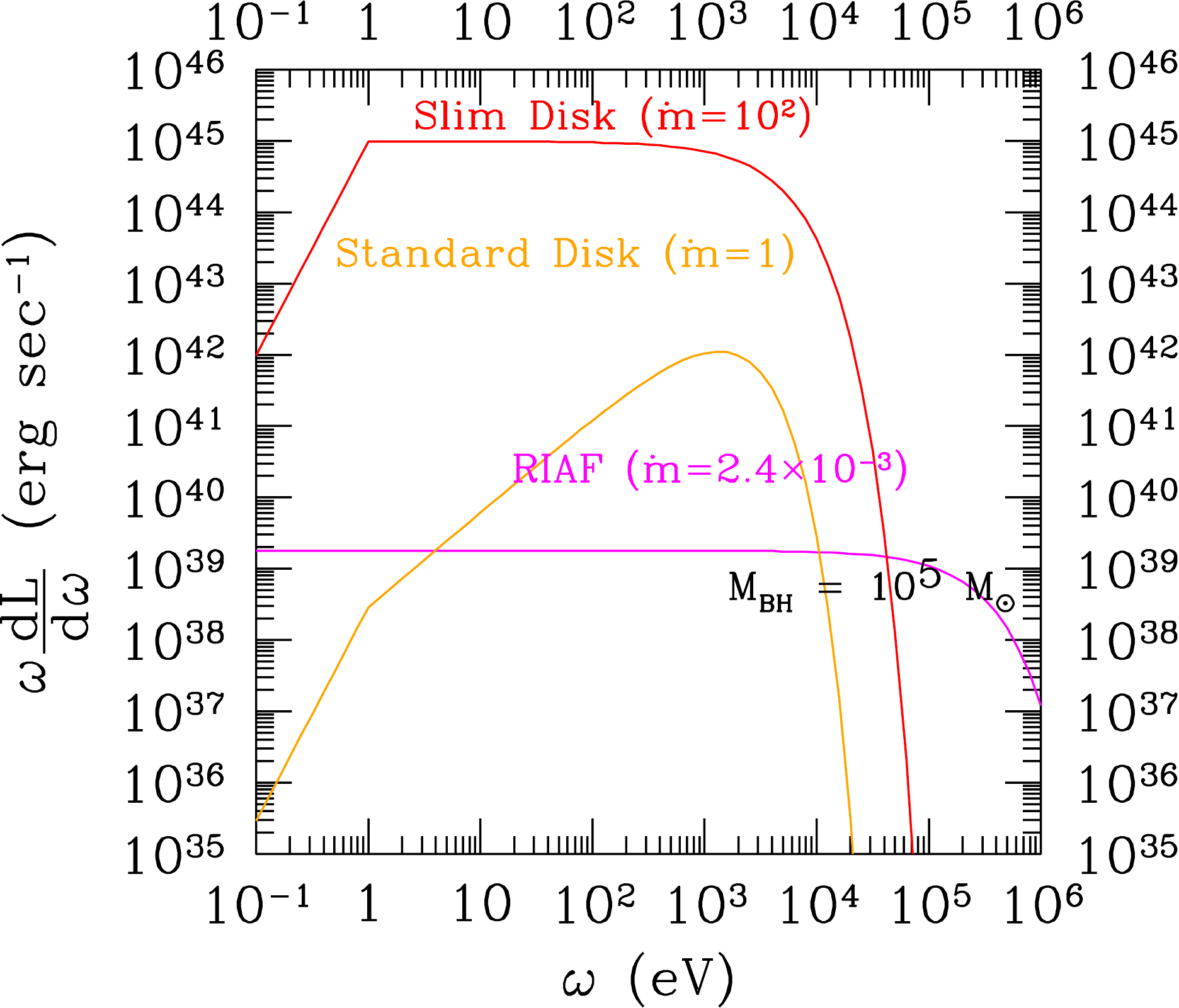}
\caption{Plots of $\omega \frac{dL}{d\omega}$ as a function of the photon energy $\omega$ in eV for $M_{\rm BH}$ = 10 $M_{\odot}$ (left) and $M_{\rm BH}$ = $10^5 M_{\odot}$ (right).}
\label{fig:nuLnu}
\end{figure}

The spectrum of the emissivity in each model is approximately parametrized
by the following function form,
\begin{eqnarray}
  \label{eq:parametrization}
  \frac{dL}{d\omega} = A_{\rm norm} 
\left (\frac{\omega}{\omega_{\rm min}} \right)^{p_s}
\exp\left[ - \frac{\omega}{\omega_{\rm cut}}\right],
\end{eqnarray}
with the photon energy $\omega$, $A_{\rm norm}$ being the normalization factor in unit of erg sec$^{-1}$ eV$^{-1}$ to be consistent with the curve plotted in Fig.~\ref{fig:lm}. These parameters are fitted to be the values shown in Table~I~(See Refs.~\cite{Watarai:2000,Yuan:2014gma,Mineshige:2008}).
\begin{table}[ht]
    \centering
\begin{tabular}{c|c|c|c}
\hline \hline & RIAF & Standard disk & Slim disk \\
\hline $\omega_{\rm min} $ in eV & $\left( M_{\rm BH}/10 M_{\odot} \right)^{-1/2}$ & 1 & 1 \\
\hline $p_s (\omega \ge \omega_{\rm min})$ & -1 & 1/3 & -1 \\
\hline $p_s (\omega < \omega_{\rm min})$ & 2 & 2 & 2 \\
\hline $\omega_{\rm cut}$  in keV & 200 & 10$\dot{m}^{1/4} \left( M_{\rm BH}/10 M_{\odot} \right)^{-1/4}$ & 10$\dot{m}^{1/4} \left( M_{\rm BH}/10 M_{\odot} \right)^{-1/4}$ \\
\hline
\hline
\end{tabular}
\label{table:params1}
\caption{
  A set of parameters parametrized in an analytical form in Eq.~(\ref{eq:parametrization})
}
\end{table}
In Fig.~\ref{fig:nuLnu}, we plot $\omega \frac{dL}{d\omega}$ as a
function of the energy $\omega$ in eV for $M_{\rm BH}$ = 10
$M_{\odot}$ (left) and $M_{\rm BH}$ = $10^5 M_{\odot}$
(right).~\footnote{See also Ref.~\cite{Kawanaka:2020uen} for possible
  modifications of the spectrum for the slim disk through the Inverse
  Compton scattering by energetic electron in coronae. In order to
  observationally confirm the shape of the spectra, we also have to
  additionally consider possible absorptions of soft X-rays by Compton
  absorbers. It is interesting that we can trace such absorbers by
  measuring neutral hydrogen with column density
  $N_{\rm HI} \sim 10^{21} {\rm cm}^{-2}$ by future low-redshifted 21
  cm observations~\cite{Moss:2017,Ursini:2017ixb,Curran:2018,Hickox:2018xjf,Morganti:2018,Liszt:2020}.}

\begin{figure}[ht]
\centering
\includegraphics[width=10cm]{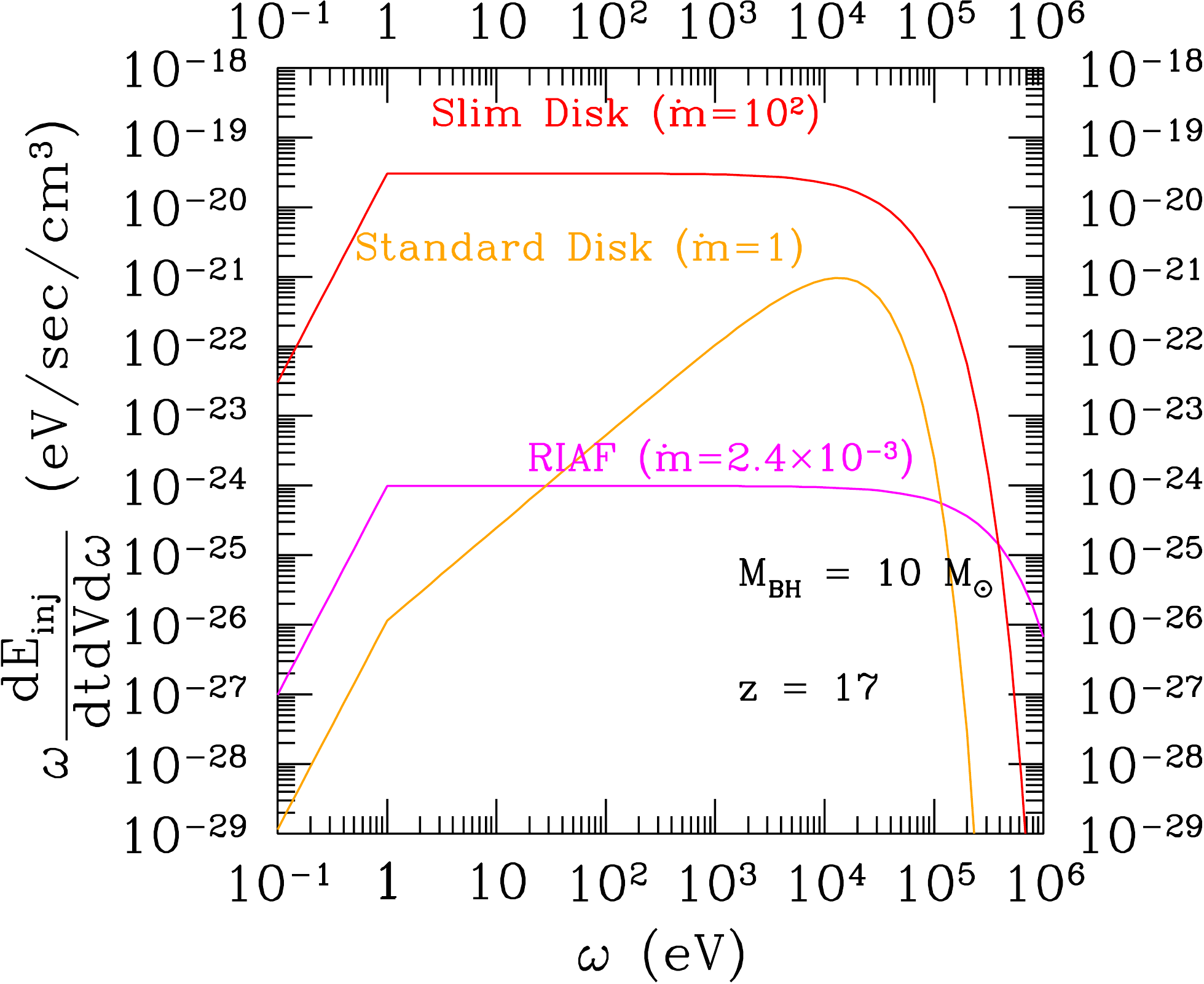}
\caption{Plots of $\omega \frac{dE_{\rm inj}}{dtdVd\omega}$ as a function of the photon energy $\omega$ in eV at $z$ = 17. Here we took the comoving number density of the seed,  $n_{\rm seed, 0}=  10^{-3} {\rm Mpc}^{-3}$ and $M_{\rm BH} = 10 M_{\odot}$.
}
\label{fig:dEinjdtdV}
\end{figure}

\section{Energy injection and deposition into the IGM from accretion disks}
\label{sec:injection}
The energy injection rate is given as
\begin{eqnarray}
  \label{eq:dEdVdt}
\frac{dE_{\rm inj}}{dVdt} (z) = \int_0^{\infty} d \omega \frac{dE_{\rm inj}}{dVdtd \omega} (z),
\end{eqnarray}
with
\begin{eqnarray}
  \label{eq:dEdVdtdomega}
\frac{dE_{\rm inj}}{dVdtd \omega} (z)  = n_{\rm seed}(z) \frac{dL}{d\omega},
\end{eqnarray}
where 
$n_{\rm seed}(z)$ is the number density of the seed BHs as a function of $z$, which are expected to be evolved to SMBHs in a late epoch. Then, we parametrize the form of $n_{\rm seed}(z)$ to be
\begin{eqnarray}
  \label{eq:nseed}
  n_{\rm seed}(z) = n_{\rm seed, 0}
(1 +z)^3 
\end{eqnarray}
with the comoving number density of the seed BHs $n_{\rm seed, 0}$.
This equation requires careful attention to the actual meaning of
$n_{\rm seed, 0}$. Surely it is equal to $n_{\rm seed}(z=0)$ at face
value. However, it is notable that $n_{\rm seed, 0}$ can increase at a
later time as a function of cosmic time, depending on models, e.g.,
for $z \lesssim 7$.  For each value of $n_{\rm seed, 0}$ thus, we take
it to be constant at least from $z \sim 30$ to $z\sim 10$. Here we may
take a maximum value of it possibly to be
$n_{\rm seed, 0} \sim {\cal O}(1) \times 10^{-3}~{\rm Mpc}^{-3}
({\Omega_{\rm CDM}h^2}/{0.1}) ({M_{\rm gal}}/{10^{12}
  M_{\odot}})^{-1}$
which is derived roughly by assuming that every (massive) galaxy at
least had a seed BH in its center in the comoving coordinate
($\sim \rho_{\rm CDM}/ M_{\rm gal}$)~\cite{Serpico:2020ehh}, with the
energy density of cold dark matter (CDM) $\rho_{\rm CDM}$, the reduced
Hubble constant $h~(\sim 0.7)$, and $M_{\rm gal}$ being a typical mass
of a massive galaxy ($\sim {\cal O}(1) \times 10^{12}
M_{\odot}$).
This value of
$n_{\rm seed, 0} \sim {\cal O}(1) \times 10^{-3} {\rm Mpc}^{-3}$ is
consistent with the observations of the SMBHs at
$z=0$~\cite{Williot:2010waa,Tanaka:2015sba}.  On the other hand, as a
conservative limit of it, we may take
$n_{\rm seed, 0} \sim {\cal O}(1) \times 10^{-7}~{\rm Mpc}^{-3}$ to
fit some observations of the SMBHs at around
$z=6$~\cite{Williot:2010waa}.~\footnote{In this case, we assume that
  the number density of the seed BHs does not increase much from $z=7$
  to $z=6$.} In this latter case, it is interpreted that the main
components of the seeds were produced at a late time $z < 6$.  Because
we cannot judge which value of the normalization of $n_{\rm seed}(z)$
is more correct, in the current study therefore, we adopt some
representative values by changing it in the range of
$n_{\rm seed, 0} = 10^{-7}~{\rm Mpc}^{-3} - 10^{-3}~{\rm Mpc}^{-3}$ as
an initial value of the comoving number density set at a higher
redshift $z \gg 17 - 20$ and study the effect in each case.  In
Fig.~\ref{fig:dEinjdtdV}, we plot
$\omega \frac{dE_{\rm inj}}{dtdVd\omega}$ as a function of the energy
$\omega$ in eV at $z = 17$. Here we took
$n_{\rm seed, 0} = 10^{-3} {\rm Mpc}^{-3}$ and $M_{\rm BH}$ = 10
$M_{\odot}$ as a reference.

To compute the deposition fractions $\{f_c(t)\}$, we need detailed information about both the spectrum of photon emitted from accretion disks and processes of their interactions with the IGM. The authors of Refs.~\cite{Shull:1985,Chen:2003gz,Padmanabhan:2005es,Ripamonti:2006gq,Kanzaki:2008qb,
  Slatyer:2009yq,Kanzaki:2009hf,Evoli:2012zz} studied how those energetic photons lose their energy through interaction processes with the IGM and affect ionization and heating of the IGM. A typical timescale of energy-loss processes for photons with their energy ranges $ 10^3\eV \simeq \omega \simeq 10^{11}\eV$ can be longer than the Hubble time. This requires detailed computation of the energy deposition fully over cosmological time scales. Here we adopt ways of computations done in Ref.~\cite{Slatyer:2015kla}\footnote{\url{https://faun.rc.fas.harvard.edu/epsilon/}}, in which the effects of energy injection is treated at a linear level, i.e, omitting higher-order nonlinear terms. For full treatments including feedback of the modification of the IGM evolution in the computation of $\{f_c(t)\}$, we refer to Ref.~\cite{Liu:2019bbm}.

By using Eqs.(\ref{eq:dEdVdt}), (\ref{eq:dEdVdtdomega}) and (\ref{eq:nseed}), we can estimate the energy injection rate analytically to be
\begin{equation}
\begin{aligned} \frac{dE_{\rm inj}}{dVdt} \sim 10^{-20}~\mathrm{eV}~\mathrm{sec}^{-1}  \mathrm{cm}^{-3} 
\times  
\left(\frac{n_{\rm seed, 0}}{10^{-3} {\rm Mpc}^{-3}}\right)^{}
\left(\frac{1+z}{18}\right)^{3}
\left(\frac{  L }{10^{40} \mathrm{erg}~\mathrm{sec}^{-1}}\right)
. 
\end{aligned}
\end{equation}
It has been known that this order-of-magnitude energy injection rate ($\sim dE_{\rm inj}/dVdt \sim {\cal O}(10^{-20})~\mathrm{eV}~\mathrm{sec}^{-1}~\mathrm{cm}^{-3}$) should have affected the absorption feature of the global 21cm line spectrum at around $z \sim 17$~\cite{Hiroshima:2021bxn}.~\footnote{  see also Refs.\cite{Poulin:2016anj,DAmico:2018sxd,Mena:2019nhm,Liu:2020wqz,Bolliet:2020ofj}.}
This means that we see intuitively that the (super-)Eddington accretion rate [$\gtrsim 10^{38} {\rm erg/sec} (M_{\rm BH}/M_{\odot}) $] can be highly constrained for $M_{\rm BH} \gtrsim 10^2 M_{\odot}$ in case of $n_{\rm seed,0}\sim 10^{-3} {\rm Mpc}^{-3}$ by observational data of the cosmological 21cm line absorption.

The time evolution of $M_{\rm BH} = M(t)$ is solved to be
\begin{eqnarray}
  \label{eq:MEdd}
  M_{\rm BH}(t) = M_{\rm BH,ini}  \exp 
\left(
\dot{m} f_{\rm duty} 
\frac{1-\eta_{\rm eff}}{\eta_{\rm eff}} \frac{t-t_{\rm ini}}{\tau_{E}} 
\right),
\end{eqnarray}
where $M_{\rm BH,ini} \equiv M_{\rm BH}(t=t_{\rm ini})$ at $t = t_{\rm ini}$ with a constant $\dot{m}$ and a possible suppression factor $f_{\rm duty} \lesssim 1$~\cite{Sassano:2021maj,Pacucci:2021ubg} due to efficiencies for a continuous accretion and so on. We adopt $f_{\rm duty} = 1$ as a simple reference value in this study. Here the timescale of the Eddington accretion is given by
\begin{eqnarray}
  \label{eq:tauE}
\tau_{E} \equiv \frac{M_{\rm BH} c^{2}}{L_{E}}=\frac{\sigma_{T} c}{4 \pi \mu G m_{p}} \simeq 0.45 \mathrm{Gyr}.
\end{eqnarray}

\section{Results}
\label{sec:results}

\begin{figure}[ht]
\centering
\begin{tabular}{cc}
\includegraphics[width=9cm]{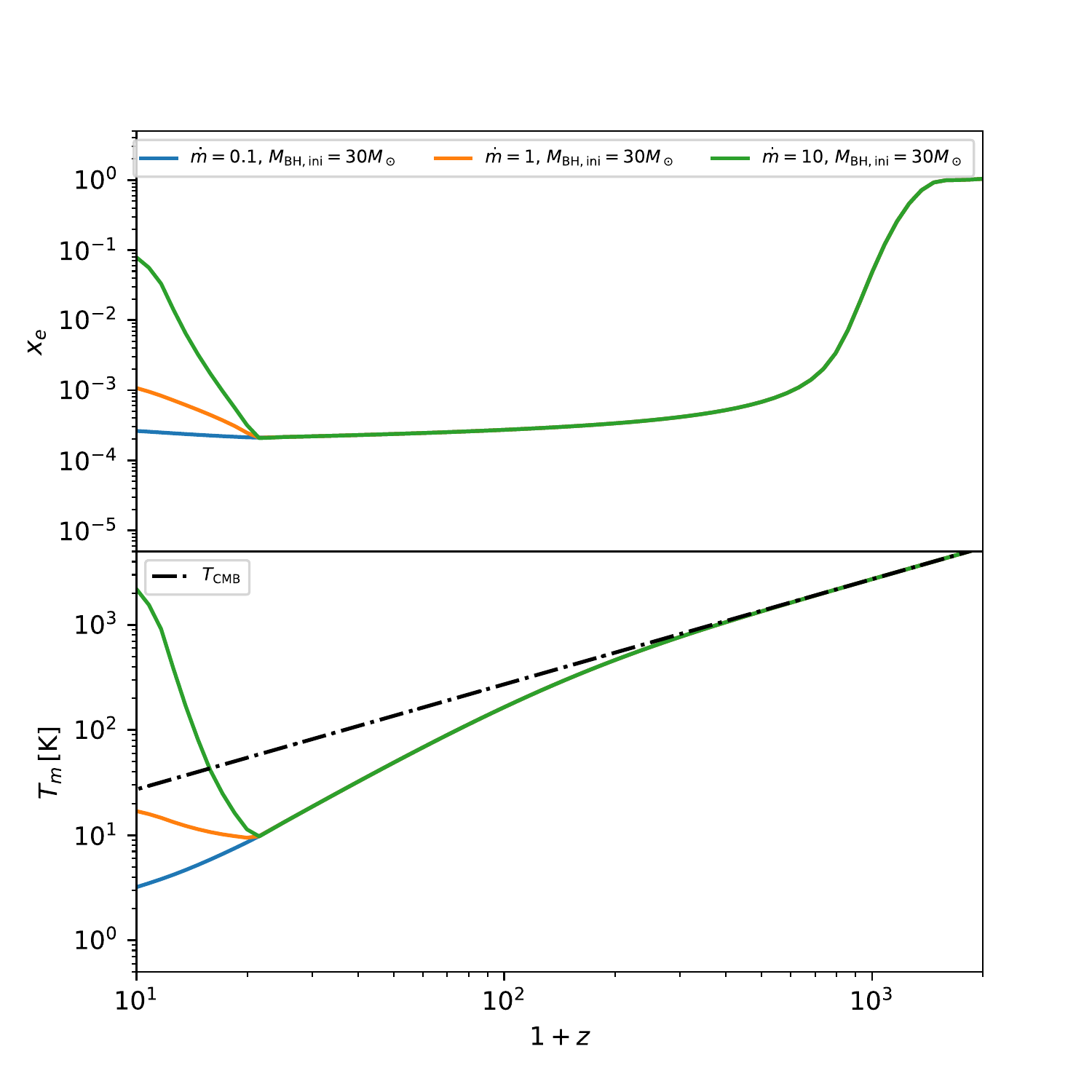} &
\includegraphics[width=9cm]{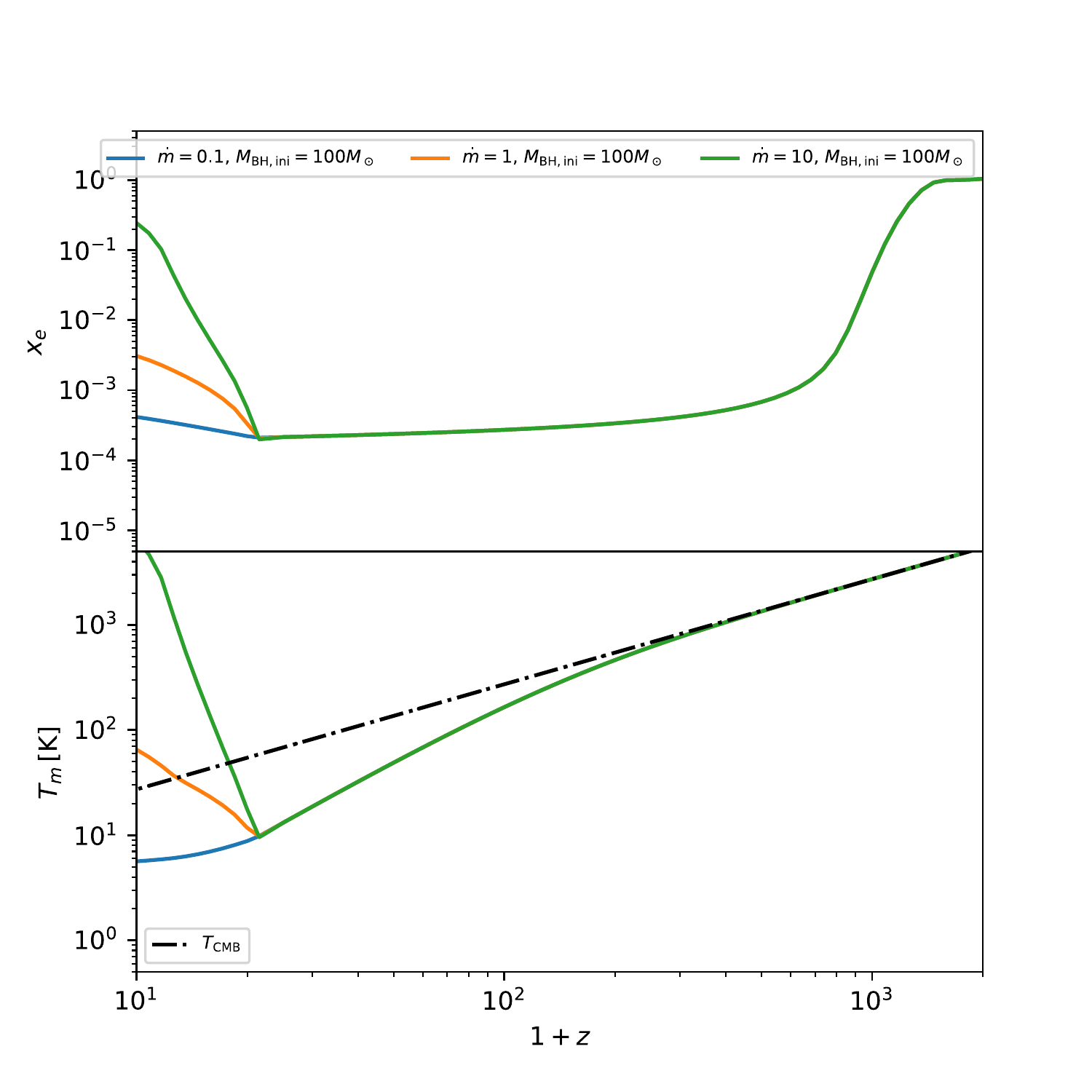} \\
\end{tabular}
\caption{\label{fig:TempEvolv} Evolutions of the ionization fraction  $x_e$ and the temperatures of gas $T_m$ in the heating by photons emitted from accretion disks with the initial redshift $z_{\rm ini} = 30$. We   plot the cases of the initial black hole mass for $M_{\rm BH, ini} = 30 M_{\odot}$ (left) and $100 M_{\odot}$ (right), respectively. The normalized mass-accretion rate is taken to be $\dot{m}$ = $0.1$ (blue), $1$ (orange), and $10$ (green).  For reference, the photon temperature $T_\gamma(z)$ (black dashed) is also plotted in each panel.}
\end{figure}

In Fig.~\ref{fig:TempEvolv}, the evolutions of $x_e(z)$ and $T_m(z)$ are shown as a function of redshift $z$ in case with the emission from accretion disk, which started from the initial redshift $z_{\rm ini} = 30$. Here we assumed there is no significant heating from the other astrophysical sources.

Compared to cases where such an energy injection is absent, the gas temperatures $T_m$ is highly enhanced. This is due to the extra heating by photons emitted from the accretion disks, which modified the evolution of the spin temperature, $T_s$, associated with the hyperfine splitting in the ground states of neutral hydrogen. This allows us to constrain emission from accretion disks from observations of differential brightness temperature of redshifted 21~cm line emission $T_{\rm 21cm}$ before reionization (See e.g. \cite{Furlanetto:2006jb}),
\begin{eqnarray}
T_{\rm  21cm}(z)=\frac{T_s(z)-T_\gamma(z)}{1+z} \tau_{\rm 21cm}(z).
\end{eqnarray}

\begin{figure}[ht]
\centering
\includegraphics[width=8cm]{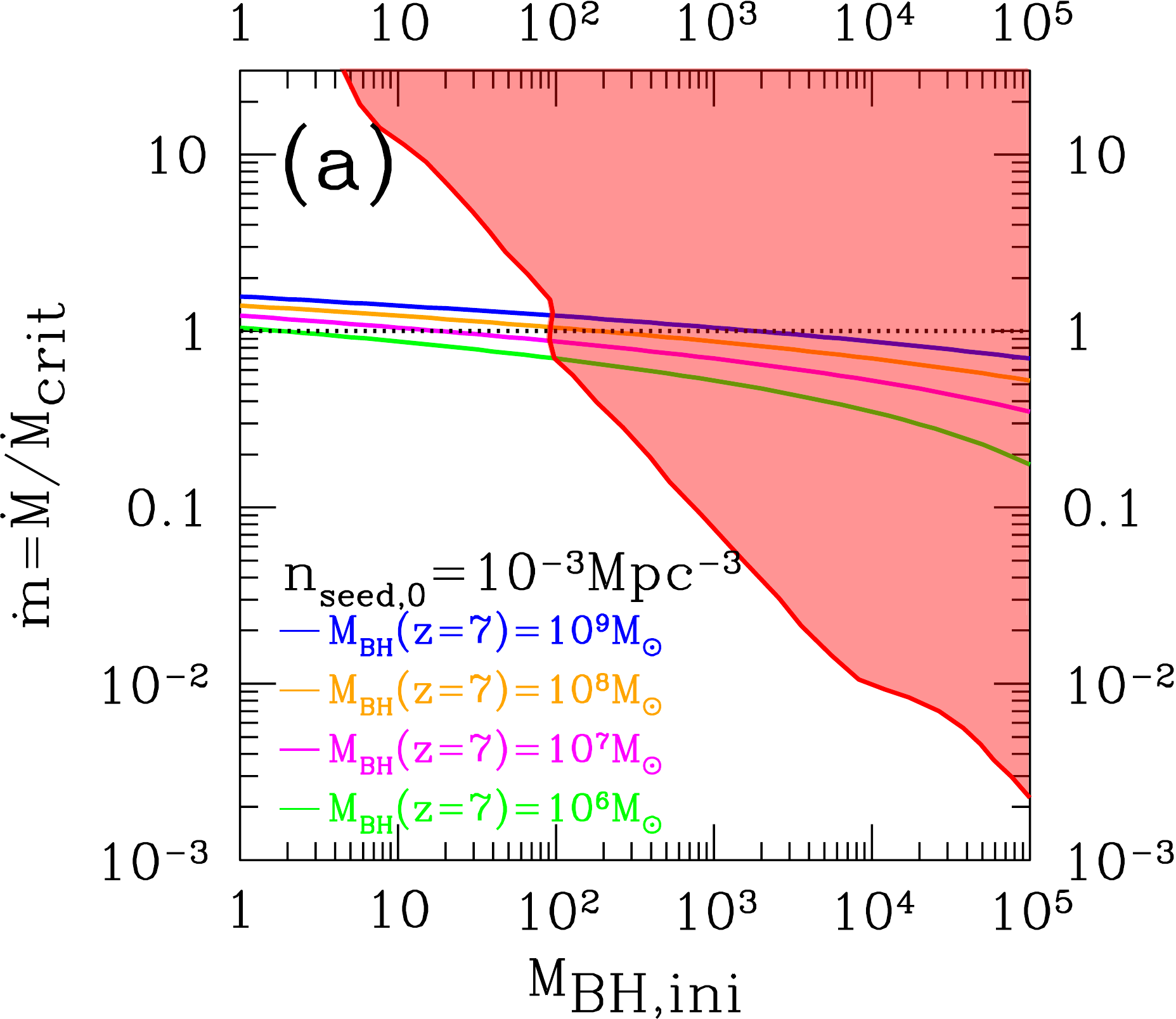}
\includegraphics[width=8cm]{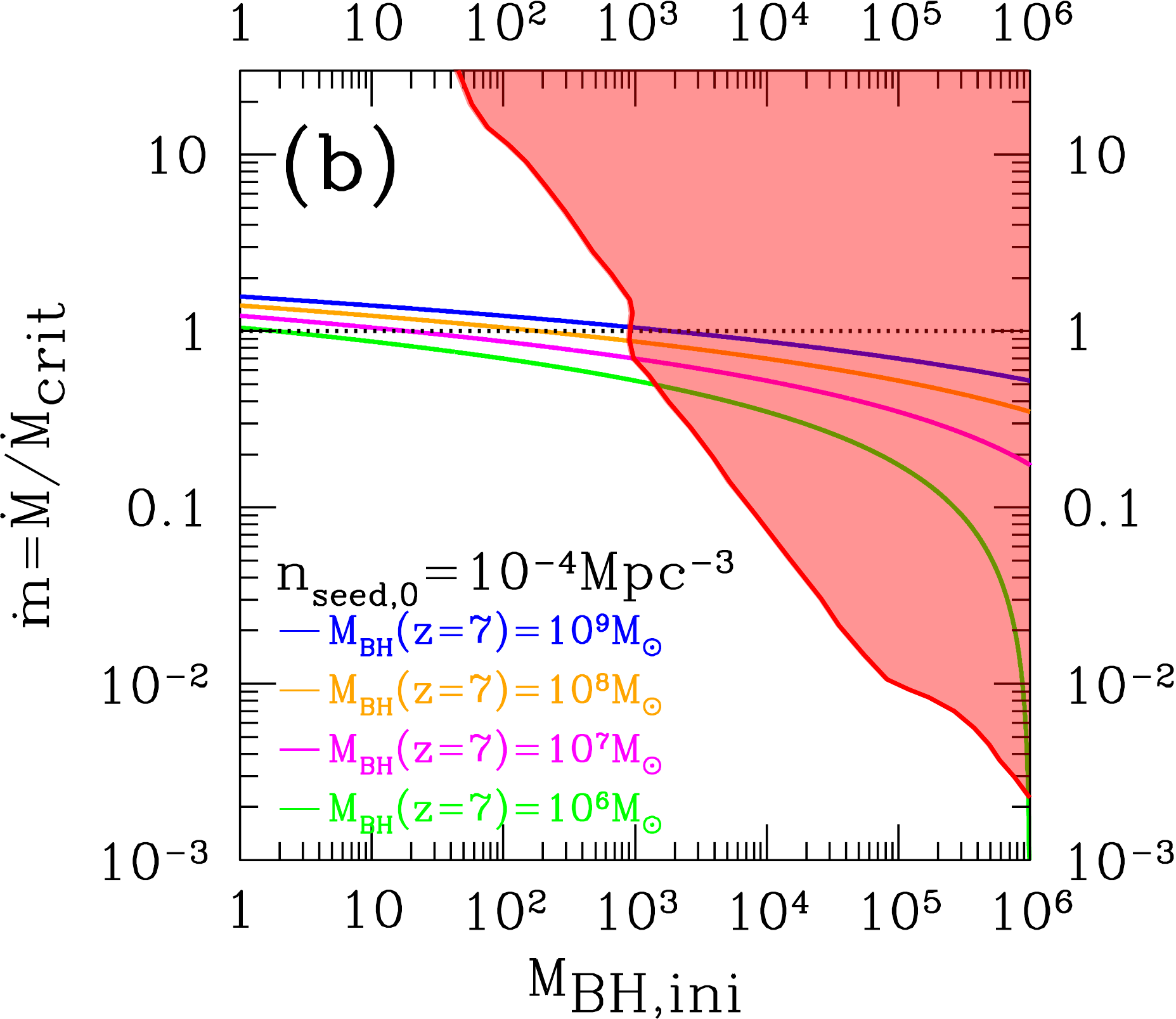}
\includegraphics[width=8cm]{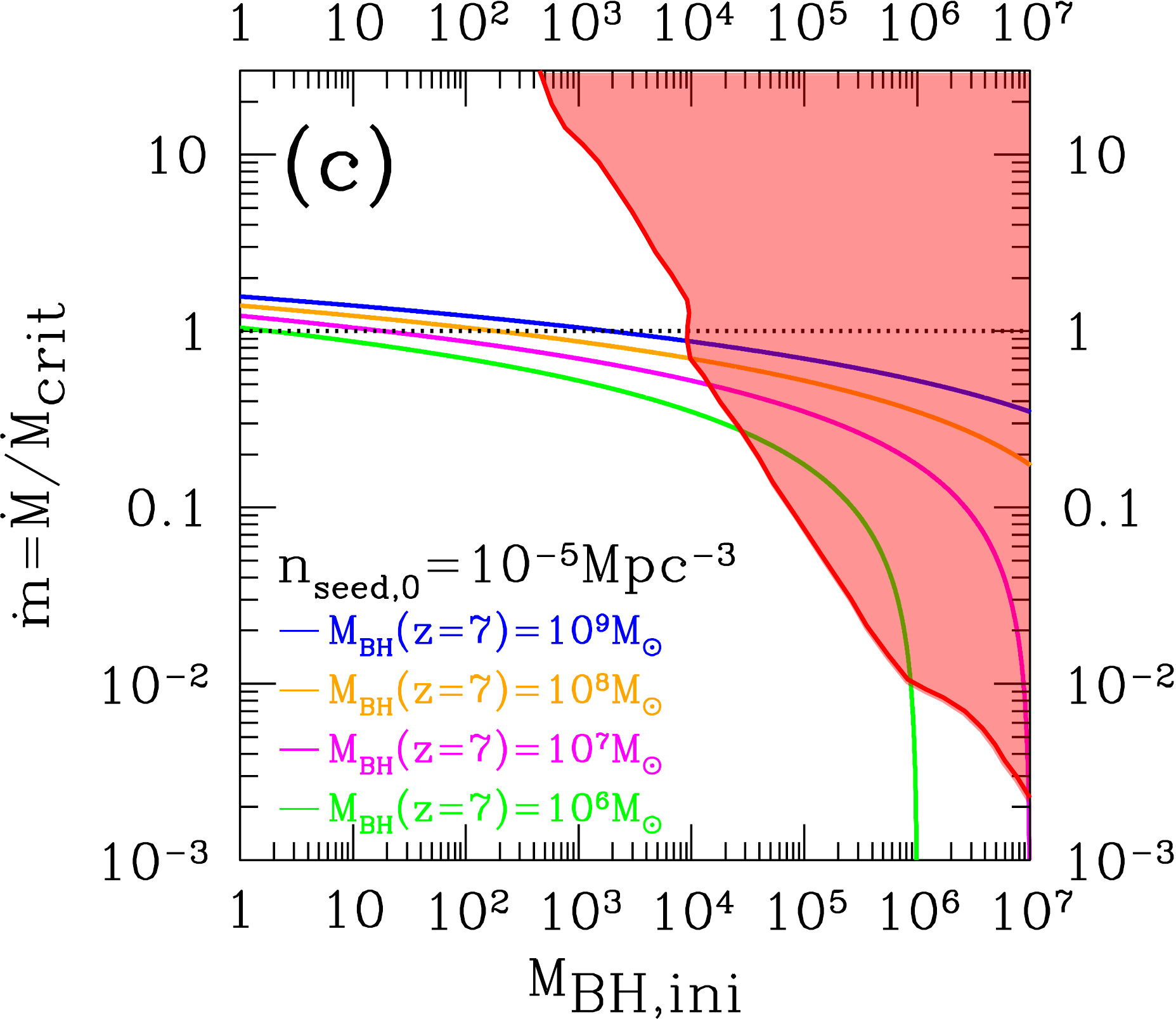}
\includegraphics[width=8cm]{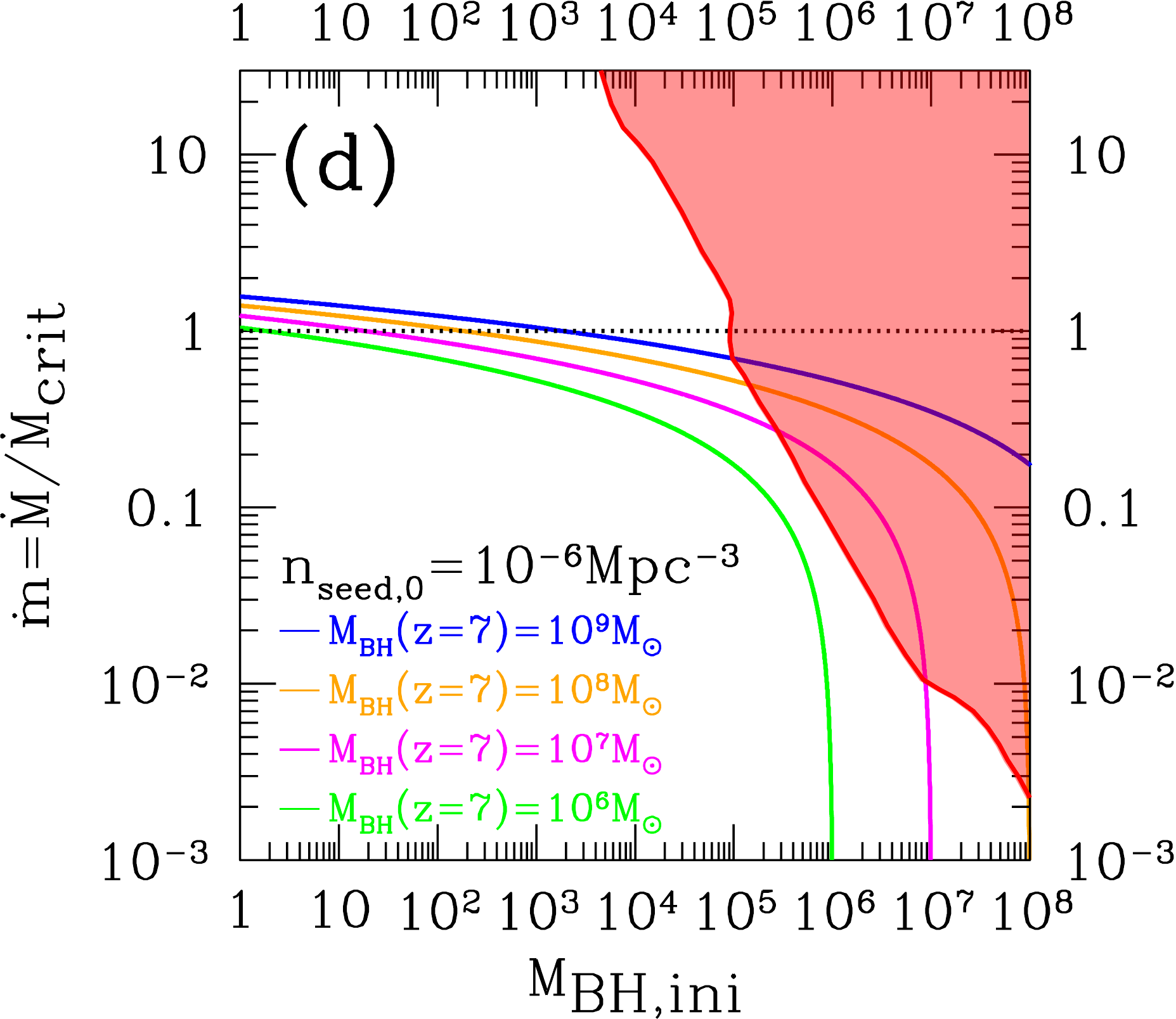}
\caption{Upper bounds on accretion rates normalized by the Eddington accretion rate as a function of initial black hole masses set at the initial redshift $z_{\rm ini}$=30.  The shaded region is observationally excluded  for $n_{\rm seed,0}/{\rm Mpc}^{-3}$ chosen to be (a) $10^{-3}$, (b) $10^{-4}$, (c) $10^{-5}$, and (d) $10^{-6}$. The blue, orange, magenta, and green solid lines denote the conditions given in Eq.~(\ref{eq:MEdd}), on which successfully $M_{\rm BH}/M_{\odot} = 10^9$,$10^8$, $10^7$, and $10^6$ are realized at $z=7$ with $f_{\rm duty} = 1$ from top to bottom, respectively.
}
\label{fig:constraints3456}
\end{figure}

\begin{figure}[ht]
\centering
\includegraphics[width=12cm]{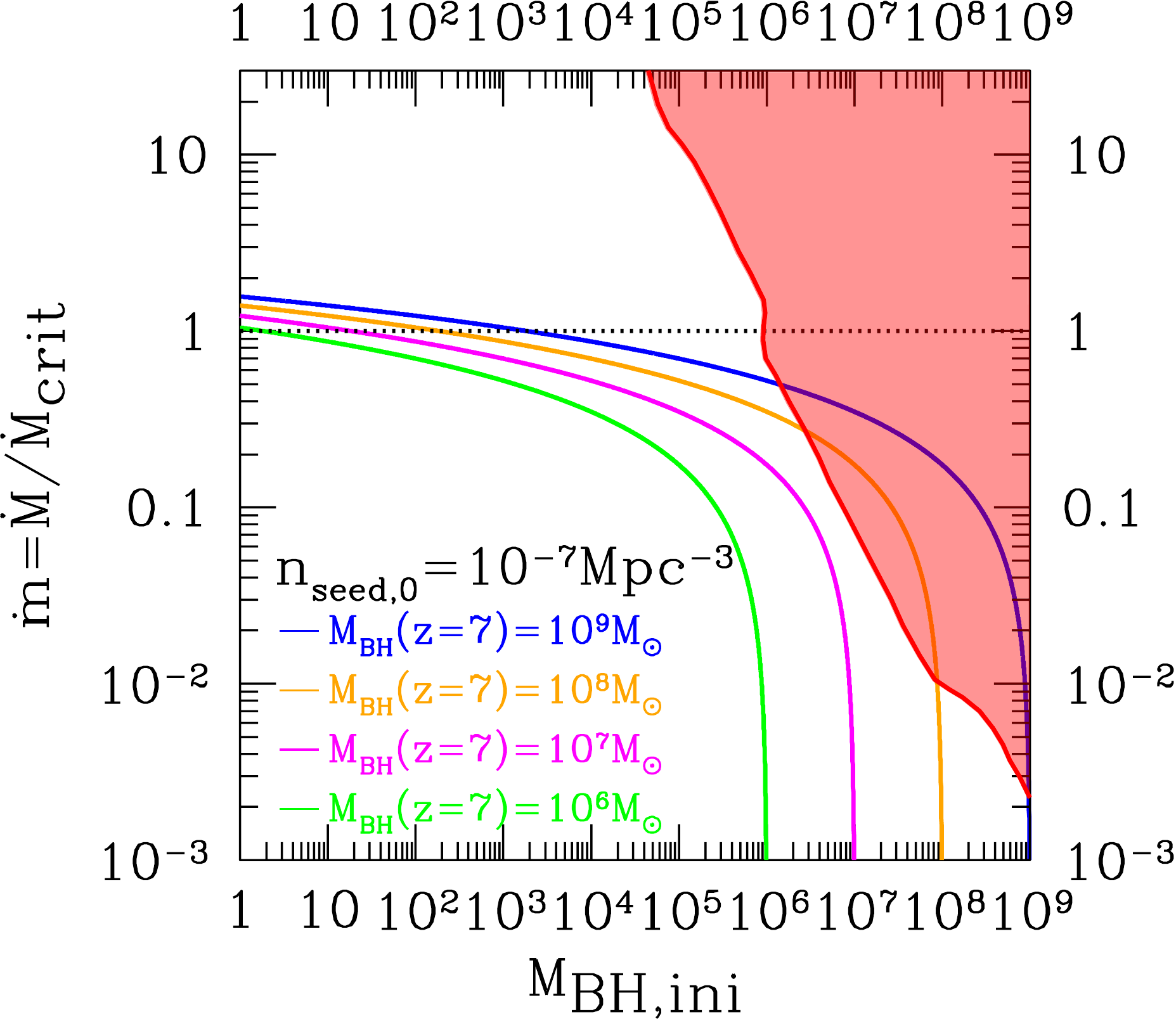}
\caption{Upper bounds on accretion rates normalized by the Eddington accretion rate as a function of initial black hole masses set at the initial redshift $z_{\rm ini}$=30. The notations are the same as those in Fig.~\ref{fig:constraints3456} but for $n_{\rm seed,0}/{\rm Mpc}^{-3}=10^{-7}$.
}
\label{fig:constraints7}
\end{figure}

A time-evolution of the spin temperature $T_s$ is controlled by relative couplings of $T_s$ with the photon temperature $T_\gamma$, the gas temperature $T_m$, and the color temperature $T_c$, which is the effective temperature associated with background Lyman-$\alpha$ radiation.  Throughout the epoch we are currently studying, the IGM is fully optically thick for emissions of Lyman-$\alpha$ radiation. Therefore, it is reasonable to assume $T_c\approx T_m$. The EDGES collaboration reported a global absorption (not emission) signal~\cite{Bowman:2018yin}
\begin{eqnarray}
 T_{\rm  21cm}=- 500^{+200}_{-500}\quad {\rm mK}\quad (99\% {\rm CL}), 
\end{eqnarray}
which means the gas temperature is smaller than the photon temperature
$T_m<T_\gamma$. Here we considered the delayed deposition which was
studied in Ref.~\cite{Liu:2018uzy,Basu:2020qoe}.~\footnote {The
  optical depth for the X-ray with a few keV highly depends on
  redshifts with a rapidly-changing function of $z$, and accidentally becomes
  {\cal O}(1) at $z$ = 10 -- 30. This is clearly shown in literature,
  e.g., Fig.2 of Ref.~\cite{Chen:2003gz} and Fig.3 of
  Ref.~\cite{Slatyer:2015kla}. Therefore, to be thermalized for the
  emitted X-rays, we need a time of the order of or even much longer than
  the Hubble time at that time. That is the reason why the
  thermalization was not realized immediately and got delayed.}

In general, photon emissions from accretion disks suppress the
amplitude of the absorption for the global 21cm signals by ionizing
and heating the IGM. It is reasonable to assume a tight coupling of
spin temperature to gas temperature through the Lyman-$\alpha$ pumping
(the Wouthuysen-Field effect \cite{Wouthuysen:1952,Field:1959}). This
maximizes the absorption depth, which gives the most conservative
limit on extra-photon emissions.  In addition, we did not assume other
ambiguous astrophysical heating sources such as UV emissions from
stars formed by non-standard CDM halo formations at small scales,
annihilating/decaying dark matter, and so on, that also helps to
obtain the most conservative upper bound on it. The prediction in the
standard $\Lambda$CDM model without such a heating by the accretion
disks gives $T_{\rm 21cm}\simeq - 230\,{\rm mK}$ (e.g., see
Ref.~\cite{Hiroshima:2021bxn}).  We obtain an upper bound on the
photon emissions from accretion rate by requiring
$T_{\rm 21cm} \le -75\,{\rm mK}$, which correspond to the 2$\sigma$
upper bound on it with given uncertainties of the EDGES
($\Delta T_{\rm 21cm} \simeq + 155$~mK) at 95$\%$
C.L.~\cite{Hiroshima:2021bxn} Here we did not assume any exotic
cooling mechanisms such as interaction between baryon and CDM only to
fit the observational depth of the absorption feature reported by
EDGES (see also~\cite{DAmico:2018sxd}). ~\footnote{Quite recently the
  SARAS 3 collaboration reported that they rejected the signal by the
  EDGES at 95.3 $\%$ C.L.~\cite{Singh:2021mxo}.  In this paper
  however, we only use the sensitivity on the errors (not the signal
  of the absorption feature) of the EDGES, which is not inconsistent
  with the claim by the SARAS 3. See similar discussions, e.g., in
  Ref.~\cite{Saha:2021pqf}. }

In Fig.~\ref{fig:constraints3456}, we show the upper bound on the accretion rate as a function of the initial seed BH mass ($M_{\rm BH, ini}$) conservatively-obtained in this study, which means that the shaded region (the upper-right region) is excluded for $n_{\rm seed,0}$ chosen to be (a) $10^{-3} {\rm Mpc}^{-3}$, (b) $10^{-4} {\rm Mpc}^{-3}$, (c) $10^{-5} {\rm Mpc}^{-3}$, and (d) $10^{-6} {\rm Mpc}^{-3}$. The blue, orange, magenta, and green solid lines denote the conditions given in Eq.~(\ref{eq:MEdd}), on which successfully $M_{\rm BH}=10^9 M_{\odot} $, $10^8 M_{\odot}$, $10^7 M_{\odot}$, and $10^6 M_{\odot}$ are realized at $z=7$ with $f_{\rm duty} = 1$ and $\eta_{\rm eff}^{-1} = 10$, respectively.  In case of $f_{\rm duty} < 1$ or $\eta_{\rm eff}^{-1} \ne 10$, readers can read off a value of the $y$-axis by using the scaling law of it $\left[\propto \dot{m} f_{\rm duty} (\eta_{\rm eff}^{-1}-1)\right]$.~\footnote{We do not use the data reported by the HERA Phase 1 to obtain the mild lower bound on 21cm emissions at around $z \lesssim 10$.~\cite{HERA:2021bsv} because it does not constrain any model parameter in the current setup where we have not specified when the standard processes of the cosmological reionization occurred.}
Here we have chosen the reference value of the initial redshift to be $z_{\rm ini} = 30$ after that the accretion had started by following the discussions in~\cite{Bromm:2013iya}. Even if we adopted a larger value of $z_{\rm ini}$, the obtained bound becomes much stronger. Thus, our current choice gives a conservative bounds on the plane of ($M_{\rm BH,~ini}$, $\dot{m}$). A more comprehensive analysis by studying every case with changing those values is outside the scope of the current paper.

In case of the maximum value of $n_{\rm seed,0} = 10^{-3} {\rm Mpc}^{-3}$ shown in Fig.~\ref{fig:constraints3456}(a), to satisfy the condition for the successful SMBH formation\ with $M_{\rm BH}=10^9 M_{\odot}$ at $z=7$, the initial masses of seed BHs have been excluded for $M_{\rm BH,~ini} \gtrsim 10^2 M_{\odot}$ at 95$\%$ C.L.  Therefore, the reference value of the initial seed mass $M_{\rm BH,~ini} \sim 10^{3.3} M_{\odot}$ with the just-on Eddington accretion rate is apparently excluded in this case. This means that we inevitably need the super-Eddington accretion rate for lighter-mass seed BHs, $M_{\rm BH,~ini} \lesssim 10^2 M_{\odot}$.

In Fig.~\ref{fig:constraints7}, we plot the excluded region when we adopt the most conservative case, $n_{\rm seed,0} =10^{-7} {\rm Mpc}^{-3}$. From this figure, we find that the initial mass of the seed BHs for $M_{\rm BH,~ini} \lesssim 10^6 M_{\odot}$ are allowed at 95$\%$ C.L. to obtain $M_{\rm BH}=10^9 M_{\odot}$ at $z=7$. In this case, we need $\dot{M} \sim 0.5 \dot{M}_{\rm crit}$ for continuous accretions on to the initial seed mass $M_{\rm BH,~ini} = 10^6 M_{\odot}$ set at $z_{\rm ini}=30$ until $z=7$.

From Fig.~\ref{fig:constraints3456} and Fig.~\ref{fig:constraints7}, readers can read off every constraint by changing $n_{\rm seed,0}$. For example, according to information of Fig.~8 of Ref.~\cite{Williot:2010waa}, to realize $M_{\rm BH}=10^7 M_{\odot}$ until $z=7$ with $n_{\rm seed,0}=10^{-5} {\rm Mpc}^{-3}$ we can see $M_{\rm BH,~ini} \lesssim 10^4 M_{\odot}$ are allowed at 95$\%$ C.L. from the magenta line in Fig.~\ref{fig:constraints3456}(c).

\begin{figure}[ht]
\centering
\includegraphics[width=8cm]{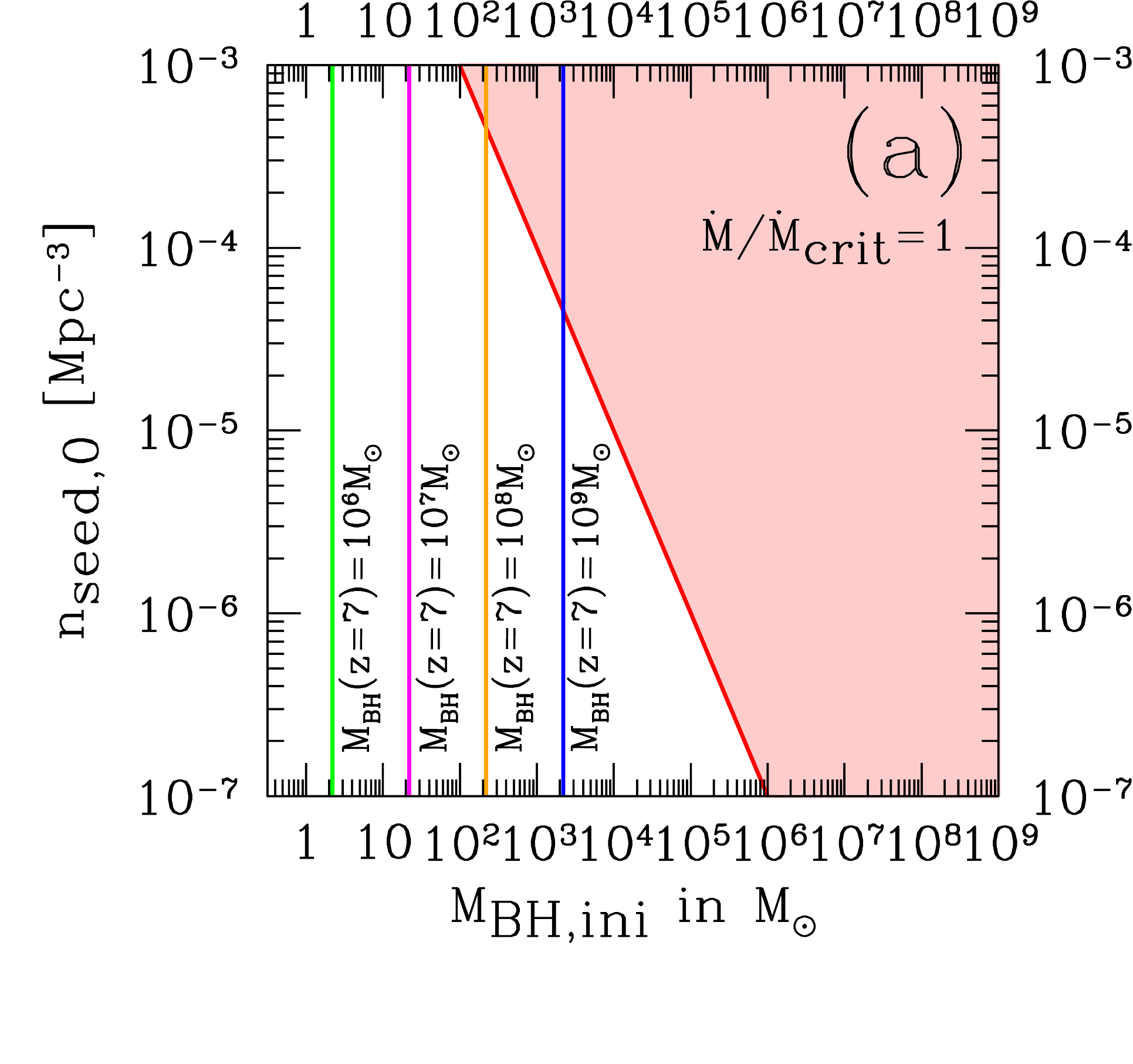}
\includegraphics[width=8cm]{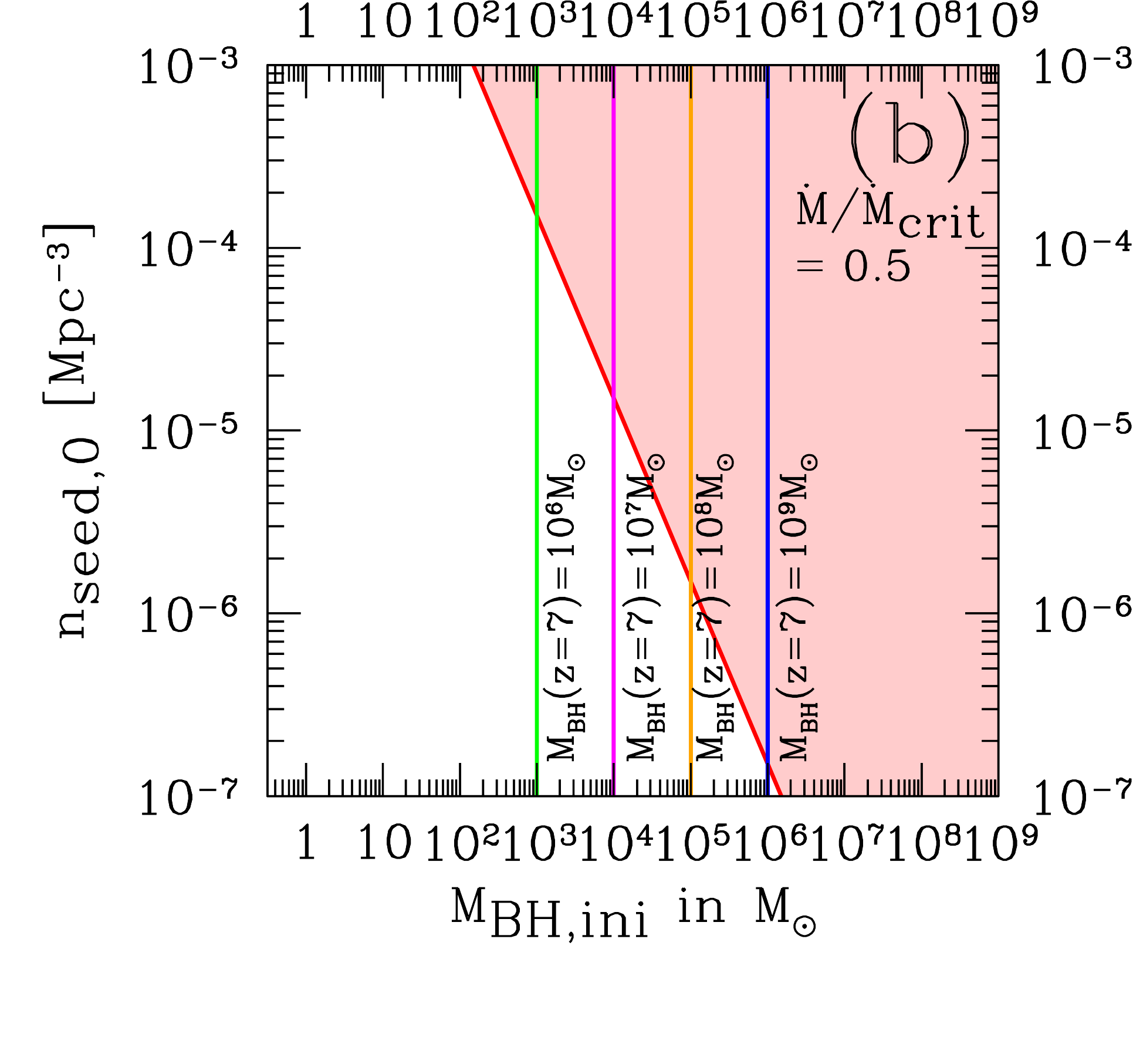}
\caption{Upper bounds on the 2D plane of the initial BH mass
  $M_{\rm BH, ini}$ set at $z_{\rm ini}=30$ and the comoving number
  density of the seed BHs $n_{\rm seed,0}$. The lines give the
  conditions to realize $M_{\rm BH}(z=7)=$ $10^9M_{\odot}$,
  $10^8M_{\odot}$, $10^7M_{\odot}$, $10^6M_{\odot}$ in case of (a)
  $\dot{M}/ \dot{M}_{\rm crit} = 1.0$, and (b)
  $\dot{M}/ \dot{M}_{\rm crit} = 0.5 $ by the accretions. The
  upper-right regions (red regions) are excluded by the observational
  bounds on the 21 cm lines at around $z\sim 17$.}
\label{fig:crossing}
\end{figure}

In Fig.~\ref{fig:crossing}, we plot the lines of the conditions to
realize $M_{\rm BH}(z=7)=$ $10^9M_{\odot}$, $10^8M_{\odot}$,
$10^7M_{\odot}$, $10^6M_{\odot}$ in the 2D plane of the initial BH
mass $M_{\rm BH, ini}$ set at $z_{\rm ini}=30$ and the comoving number
density of the seed BHs $n_{\rm seed,0}$ in case of (a)
$\dot{M}/ \dot{M}_{\rm crit} = 1.0$, and (b)
$\dot{M}/ \dot{M}_{\rm crit} = 0.5 $. The upper-right regions (red
regions) are excluded by the observational bounds on the 21 cm lines
at around $z\sim 17$. From Fig.~\ref{fig:crossing}(a), to realize
$M_{\rm BH}(z=7)=$ $10^9M_{\odot}$, we find that
$n_{\rm seed,0} \gtrsim 4 \times 10^{-5} {\rm Mpc}^{-3}$ is excluded
for $\dot{M}/ \dot{M}_{\rm crit} \gtrsim 1.0$. This means that we need
another mechanism to create the seed BHs after $z \ll 17$. On the
other hand, from Fig.~\ref{fig:crossing}(b), we obtained the
conservative upper limit on $M_{\rm BH, ini}$ to be
$10^6 M_{\odot}$ with the minimum mass-accretion rate
$\dot{M}/ \dot{M}_{\rm crit} \gtrsim 0.5$ to realize
$M_{\rm BH}(z=7)=$ $10^9M_{\odot}$ for a conservative lower limit on the
comoving number density of the seed BHs
$n_{\rm seed,0}= 10^{-7} {\rm Mpc}^{-3}$.

It is notable that those bounds can be applicable to scenarios for accretions on to primordial black holes (PBHs) like as floating BHs, which were produced at $z \gg 30$.~\footnote{We can refer to papers for the other aspects of researches about the PBHs constrained from 21cm line observations for cosmological accretions on to PBHs~\cite{Gong:2017sie,Gong:2018sos,Hektor:2018qqw,Mena:2019nhm,Hutsi:2019hlw,Hasinger:2020ptw,Villanueva-Domingo:2021cgh,Yang:2021idt,Yang:2021agk,Acharya:2022txp} and evaporation of PBHs~\cite{Mack:2008nv,Carr:2009jm,Clark:2018ghm,Carr:2020gox,Halder:2021jiv,Natwariya:2021xki,Mittal:2021egv,Cang:2021owu,Saha:2021pqf}.}
Those BHs could enter into halos by chance and be surrounded by dense gas at higher redshifts. In this case, the cosmological omega parameter of the seed BHs ($\Omega_{\rm sBH}$) is related to $n_{\rm seed,0}$ by
\begin{eqnarray}
  \label{eq:OmegaSeedBH}
\Omega_{\rm sBH} / \Omega_{\rm CDM} \sim 10^{-10} 
\left(
\frac{n_{\rm seed,0}}{10^{-3} {\rm Mpc}^{-3}}
\right) 
\left(
\frac{M_{\rm BH, ini}}{10^2 M_{\odot}}
\right) 
\left(
\frac{M_{\rm SMBH}}{10^9 M_{\odot}}
\right)
\left(
\frac{M_{\rm gal}}{10^{12} M_{\odot}}
\right)^{-1}.
\end{eqnarray}
with $M_{\rm SMBH}$ the mass of SMBHs.  Then, $M_{\rm BH, ini}$ has two meanings: 1) the mass of a BH which was originally equal to it, or 2)  which had evolved to this value by an accretion until $z=z_{\rm ini}$. This parametrization also requires careful attention to the meaning of $\Omega_{\rm sBH}$ (or $n_{\rm seed,0}$) here. It is different from the usual definition of the cosmological omega parameter for the homogeneously-distributed field component of the BHs, but for the one inside halos surrounded by rich gas. Actually it should be highly model-dependent to estimate the real fraction of such seed BHs captured into this kind of systems to the total BHs.

\section{Conclusion}
\label{sec:conclusion}

In this paper we have studied the scenarios of the accretions on to black holes from sub- to super- Eddington rates at high redshifts $z \gg 10$ which are expected to become seeds to evolve to supermassive black holes until redshift $z \sim 7$. Such an accretion disk emits copious high-energy photons (the UV and keV-MeV photons) which had heated the plasma of the intergalactic medium continuously at high redshifts. In this case, the gas temperature is modified, by which the absorption of the cosmological 21 cm lines are suppressed at around $z \sim 17$. 

As is shown in Fig.~\ref{fig:constraints3456} and Fig.~\ref{fig:constraints7}, by comparing the theoretical prediction of the global cosmological 21cm line absorption with the signal observed by the EDGES collaboration, conservatively we have obtained the upper bounds on the mass-accretion rate on to each initial seed black hole set at $z \gtrsim 20 - 30$. If we adopted a maximum value for the comoving number density of the seed BH to be $n_{\rm seed,0}=10^{-3} {\rm Mpc}^{-3}$ shown in Fig.\ref{fig:constraints7}, in order to satisfy the successful formations of the supermassive black holes until $z=7$, we obtained the upper bound on the seed-BH masses to be $M_{\rm BH, ini} \lesssim 10^2 M_{\odot}$.  Clearly the reference model $M_{\rm BH, ini} \sim 10^{3.3} M_{\odot}$ with the exact Eddington accretion rate ($\dot{m}=1$) is excluded.  In other words, for a seed BH mass smaller than $10^2 M_{\odot}$ we inevitably need the super-Eddington accretion. Alternatively, such a high accretion on to a larger seed mass ($M_{\rm BH, ini} \gtrsim 10^2 M_{\odot}$) should have started after $z \sim 17 - 20$.

On the other hand, if we adopted a conservative value for the comoving number density of the seed BH, $n_{\rm seed,0}=10^{-7} {\rm Mpc}^{-3}$ to fit the observations of SMBHs partly with $M_{\rm BH} \sim 10^9 M_{\odot}$ at $z \sim 6$~\cite{Williot:2010waa}, we obtain a milder bound on the initial seed mass, $M_{\rm BH, ini} \lesssim 10^6 M_{\odot}$. In this latter case, we only need a sub-Eddington rate for $M_{\rm BH, ini} \sim  10^{3.3} - 10^6 M_{\odot}$.

This constraint is applicable to scenarios for accretions on to primordial black holes (PBHs) and so on. Then, the cosmological omega parameter of the seed PBHs,i.e., $\Omega_{\rm sBH}$ (not the homogeneously-distributed field component of the PBHs, $\Omega_{\rm PBH}$) is related to $n_{\rm seed,0}$ approximately by $\Omega_{\rm sBH} / \Omega_{\rm CDM} \sim 10^{-10} (n_{\rm seed,0}/10^{-3} {\rm Mpc}^{-3})(M_{\rm BH, ini}/10^2 M_{\odot}) (M_{\rm SMBH}/10^9 M_{\odot})(M_{\rm gal}/10^{12} M_{\odot})^{-1}$.

In future, more precise data of high-redshifted 21cm lines will be reported by HERA~\cite{Beardsley:2014bea}, SKA~\cite{SKAspec}, Omniscope~\cite{Omniscope} or DAPPER~\cite{Burns:2021ndk}.  By adopting those data, then we will be able to detect signatures of the super-Eddington accretion on to the seed BHs to evolve to the high-redshifted SMBHs.

\acknowledgments We thank Kohei Inayoshi, Norita Kawanaka, Koutarou Kyutoku, Vivian Poulin and Pasquale D. Serpico for useful discussions. This work was supported in part by JSPS KAKENHI Grant Numbers JP17H01131 (K.K. and T.S.), JP15H02082 (T.S.), JP18H04339 (T.S.), JP18K03640 (T.S.), and MEXT KAKENHI Grant Numbers JP19H05114 (K.K.), JP20H04750 (K.K.). S.W. is partially supported by the grants from the National Natural Science Foundation of China with Grant No. 12175243, the Institute of High Energy Physics with Grant No. Y954040101, and the Key Research Program of the Chinese Academy of Sciences with Grant No. XDPB15. Numerical computations were carried out on Cray XC50 at Center for Computational Astrophysics, National Astronomical Observatory of Japan.





\end{document}